\documentclass[11pt]{article}
\usepackage{setspace,amsmath}
\usepackage{epsfig}
\begin{document}

\begin{center} 
\bf {\large {Impact of Investor's Varying Risk Aversion on the Dynamics of Asset Price Fluctuations}}
\end{center}
\begin{center} 
Baosheng Yuan and Kan Chen
\end{center}
\begin{center} 
Department of Computational Science, Faculty of Science, National University of Singapore, Singapore 117543
\end{center}
\date{today}

\begin{abstract}

While the investors' responses to price changes and their price forecasts are well accepted major factors contributing to large price fluctuations in financial markets, our study shows that investors' heterogeneous and dynamic risk aversion (DRA) preferences may play a more critical role in the dynamics of asset price fluctuations. 
We propose and study a model of an artificial stock market consisting of heterogeneous agents with DRA, and we find that DRA is the main driving force for excess price fluctuations and the associated volatility clustering. We employ a popular power utility function, $U(c,\gamma)=\frac{c^{1-\gamma}-1}{1-\gamma}$ with agent specific and time-dependent risk aversion index, $\gamma_i(t)$, and we derive an approximate formula for the demand function and   aggregate price setting equation.
The dynamics of each agent's risk aversion index, $\gamma_i(t)$ (i=1,2,...,N), is modeled by a bounded random walk with a constant variance $\delta^2$.
We show numerically that our model reproduces most of the ``stylized'' facts observed in the real data, suggesting that dynamic risk aversion is a key mechanism for the emergence of these stylized facts.

\end{abstract}

\section{Introduction}
There have been many attempts to construct models of financial markets and to understand the key statistical features of financial time series.
It remains a great challenge, however, due the inherent complexity of the financial market, to develop a parsimonious market model that can reproduce all key ``stylized'' facts observed in real financial data and provide insights into the market mechanism for the emergence of these stylized facts. One of the promising approaches is 
agent-based modeling, which has reproduced and explained the emergence of some of  the ``stylized'' facts.
The agent-based modeling provides an ideal framework for investigating the impact of investors' behaviors on the price dynamics from many different perspectives; it has become an indispensable tool for understanding the price dynamics of financial markets\footnote{LeBaron(2005) is a good recent survey on the field}. With agent-based modeling, one can model and investigate, for example, how investors make their price forecasts and how their price forecasts influence the price fluctuation (Arthur, Holland, LeBaron, Palmer, and Tayler (1997), DeLong, Shleifer, Summers and Waldmann(1990), Levy, Levy and Solomon (2000)),  how investors respond to price change (Lux and Marchesi(1999), Caldarelli, Marsli and Zhang (1997)),  and how investors form and change their market beliefs (Brock and LeBaron (1996), Barberis, Shleifer and Vishny (1998)). In this paper we use agent-based models to study 
how investors' fluctuating risk preferences affect the price dynamics. This important issue has not been fully explored.

Financial markets present many important and challenging problems.
First, the market consists of intelligent, competing and heterogeneous agents who, with different beliefs in the market, different abilities to acquire and process market information, and mutually conflicting interests, try to make investment decision for their own benefits. Second, each agent's decision depends on his estimate of price expectations of other agents who also make their own estimates; this precludes expectations being formed by deductive means and leaves inductive reasoning as the only choice (Arthur \textit{et al} 1997).  A market of agents employing inductive reasoning often exhibits irrational herding behavior (Bak, Paczuski and Shubik (1997), Cont and Bouchaud (2001)), resulting in excessive price fluctuations or sometimes market bubbles and crashes. In such market agents' sentiment and degrees of risk aversion play a critical role in determining its price dynamics. Third, agents can learn and therefore adapt their strategies dynamically to improve their performance; this exacerbates the unpredictability of the markets. 
The change in an investor's strategy or behavior can be the cause or the result of the investor's changing sentiment (Barberis and Shleifer 1998), represented by (pessimistic) under-reaction or (optimistic) over-reaction to market dynamics driven by arrival of new information and changing market macro/micro-environment. These characteristics of financial market (competing with different beliefs and conflicting interests, interdependence of price expectations, and unpredictable changes of risk aversion) may lead to the formation of so-called noise traders behaviors. DeLong \textit{et al} (1990) have found this a (behavioral) source for price to diverge significantly from the fundamental value in financial markets -- the so called ``noise trading effect''.
A good model, needless to say, must successfully address these challenges; more importantly, it must be able to produce simulation time series that can capture the key ``stylized'' empirical facts observed in real financial time series. This paper reports our efforts in constructing such a model. The key ingredient of our model is the inclusion of investors' changing degrees of risk aversion. We show that this is the main cause of excess stock price fluctuations and the associated volatility clustering.

We outline here our model of interacting heterogeneous agents. We first consider a baseline model, in which the agents use past price information to form their sets of future price expectations. The agents are adaptive as their price expectations are not based on one particular estimator but are determined by their best-performance estimators which may change from time to time.
The agents also use their erroneous stochastic beliefs (DeLong \textit{et al.}(1990)) in the market to make price adjustment on their best forecasts, which are assumed to be normally distributed. 

In our baseline model, we assume that the agents have a decreasing absolute, but constant relative, risk aversion (DARA and CRRA) utility function, $U(c) = (c^{1-\gamma}-1)/(1-\gamma)$. The existing agent based models of stock market, such as the Santa Fe artificial stock market model, typically use a constant absolute risk aversion (CARA) utility function, $U(c) = - e^{-\lambda c}$, for the price setting equation can be easily derived under such utility. As the focus of our paper is on risk aversion, we choose to use the well accepted (DARA) power utility function. Although a simple analytic formula for the demand function under this utility is not available, we have derived a general functional form and a rather good approximation for the demand function. Like the SFI market model, our baseline model market with a choice of the parameters corresponding to normal market conditions, exhibits some excess volatility, but not to the extent of the volatility observed in real markets. In addition, there is little enhancement of volatility clustering at high volatility regime, which is observed in real market data (Chen, Jayaprakash and Yuan (2005)).

By simply allowing investors to change their risk aversion attitudes,  we obtain excess volatility and volatility clustering in very good agreement with real market data. The implication is clear: dynamic risk aversion (instead of fixed constant risk aversion) is directly responsible for excess volatility and the associated clustering. 
Specifically, in our DRA model, which is built from our baseline model, all agents have the power utility functions ($(c^{1-\gamma}-1)/(1-\gamma)$) but with different and time varying risk aversion indices (degrees), $\gamma_{i,t}$, which we assume to follow an independent bounded random walk with a variance $\delta^2$. We will show that the magnitude of excess volatility is directly related to $\delta^2$. 
With such DRA our model market exhibits most of the important statistical characterization of real financial data, such as the ``stylized'' facts related to excess volatility (Mandelbrot (1963), Fama (1963), Bouchaud and Potters(2000), Mantegna and Stanley (1999), Cont (2001)) and volatility clustering (Mandelbrot (1963), Fama (1965), Engle (1982), Baillie, Bollerslev, and Mikkelsen (1996), Chou (1988), Schwert (1989), Poterba and Summers (1986), Chen \textit{et al.} (2005)).

We have also studied the impacts of the dynamic risk aversion on the market dynamics using a few other  baseline models, including the SFI market model, and we found similar results. This suggests that our results on dynamic risk aversion are rather generic. 

The next section contains a derivation of the price equation under the power utility function for risk aversion.
Section III describes our baseline model with a fixed constant risk aversion.
Section IV introduces our DRA model.
Section V reports the results from numerical simulations of our model. 
Section VI considers a DRA model built with the SFI market model as the baseline model. The last section summarizes.

\section{Demand Function and Price Setting Under the Power Utility Function}
We consider a market of $N$ heterogeneous agents who form their subjective expectations inductively and independently based on  their investment strategies. 
There are two assets, a risky stock paying a stochastic dividend with a limited supply of $N$ shares\footnote{In practice the number of shares is never the same as the number of agents; here we set the two numbers the same for the sake of convenience and setting them different does not change the results}, and a risk-free bond paying a constant interest rate, $r$, with infinite supply. 
All agents have the same form of power utility function, $U(c_t;\gamma_{i,t})=\frac{c_t ^{1-\gamma_{i,t}}-1}{1-\gamma_{i,t}}$, but they have their own time-dependent risk aversion indices denoted by $\gamma_{i,t}$.  
At each time step $t$, every agent decides how to allocate his wealth between the risk-free bond and the risky stock. 
Since the values for both the dividend payment and the stock price at the next period $t+1$ are unknown random variables, the investors can only estimate the probability of various outcomes.
Assume each agent's estimation at time $t$ of the next step's price and dividend is normally distributed with the (conditional) mean and variance, $E_{i,t}[p_{t+1}+d_{t+1}]$ and $\sigma^2_{i,t} (i=1,2,...N)$ respectively. 
It can be shown, by optimizing the total utility, that the demand of agent $i$ for holding the share of the risky stock is approximately,
\begin{equation} \label{eq:demand} 
D_{i,t} =\frac{E_{i,t}[p_{t+1} + d_{t+1}] - p_t(1+r)}{\gamma_{i,t} \sigma^2_{i,t}  p_t (1+r) }
\end{equation}
where $p_t$ is the stock price at time $t$, $\gamma_{i,t}$ is agent i's index (degree) of risk aversion, and $\sigma^2_{i,t}$ the conditional variance of price estimation.

The market clearing condition: $\sum_i^N D_{i,t+\tau} = \sum_i^N D_{i,t} = N$ can be used to determine the current market price and relate the price at time $t+\tau$, $p_{t+\tau}$ to the price at time $t$, $p_t$:
\begin{equation} \label{eq:price_t} 
p_t= \frac{\sum_i^N\frac{ E_{i,t}[p_{t+1} ~+~ d_{t+1}]}{\gamma_{i,t} \sigma^2_{i,t} (1+r)}} {N + \sum_i^N \frac{1}{\gamma_{i,t} \sigma^2_{i,t}}}.
\end{equation} 
and
\begin{equation} \label{eq:price_tau} 
p_{t+\tau} = \frac{ \sum_i^N \frac{E_{i,t+\tau}[p_{t+\tau+1}+d_{t+\tau+1}]} {\gamma_{i,t+\tau} \sigma_{i,t+\tau}^2 (1+r) } - \sum_i^N \frac{1} {\gamma_{i,t+\tau} \sigma_{i,t+\tau}^2}  }  { \sum_i^N \frac{E_{i,t}[p_{t+1}+d_{t+1}]} {\gamma_{i,t} \sigma_{i,t}^2 (1+r) } - \sum_i^N \frac{1} {\gamma_{i,t} \sigma_{i,t}^2}} p_t.
\end{equation}

It can be seen from the demand and price equations that the degree of the agent's risk aversion plays an important role. 

We now show the derivation of the above equations.
Assume at time $t$ agent $i$'s consumption is $c_{i,t}$, and he invests a portion $x$ of his current consumption in the risky asset.
His total utility function defined over the current and future values of consumption is:
\begin{equation} \label{eq:utility_total} 
U(c_{i,t},c_{i, t+1},\gamma_{i,t}) = U(c_{i,t},\gamma_{i,t}) + U(\frac{c_{i,t+1}(x)}{R_f},\gamma_{i,t}),
\end{equation} 
where the consumption at time $t+1$ can be written as
\begin{equation} \label{eq:utility_t+1} 
c_{i,t+1}(x) =c_{i,t} [(1-x)R_f + x \tilde{R}_{i,t+1}].
\end{equation} 
Here $R_f = 1+ r$ is the gross risk-free return  and $\tilde{R}_{i,t+1}$ the gross return on the risky asset.
Agent $i$ determines the amount of his investment on the risky asset, $x$, by maximizing his total utility, Eqn (\ref{eq:utility_total}). The maximization problem can be written as
\begin{equation} \label{eq:MEU1} 
\max_{x} E_t[U(c_{i,t},\gamma_{i,t}) + U(\frac{c_{i,t+1} (x)}{R_f}, \gamma_{i,t})] \equiv \max_{x} E_t[U(\frac{c_{i,t+1} (x)}{R_f}, \gamma_{i,t})],
\end{equation} 
The last equality follows because the utility $U(c_{i,t},\gamma_{i,t})$ is known at time $t$ and it does not contain $x$.

The power utility function is given by,
\begin{equation} \label{eq:utility_function} 
U(c_{i,t}; \gamma_{i,t}) =\frac{c_{i,t}^{1-\gamma_{i,t}}-1} {1-\gamma_{i,t}}
\end{equation} 
Substituting Eqn (\ref{eq:utility_t+1}) into Eqn (\ref{eq:utility_function}) and then inserting it back to Eqn (\ref{eq:MEU1}), the maximization is now given by
\begin{equation} \label{eq:MEU3} 
\max_{x} \frac{c_{i,t}^{1-\gamma_{i,t}}} {1-\gamma_{i,t}} E_t  \{ [1 + \tilde{r}_{i,t+1} x] ^{1-\gamma_{i,t}}\}
\end{equation} 
where $\tilde{r}_{i,t+1}=\frac{\tilde{R}_{i,t+1} - R_f}{R_f}$ is the present (discounted) value of the \textit{net} return at next time step $t$+1.
Assuming that agent $i$'s prediction errors of $\tilde{r}_{i,t+1}$ are (conditionally) normally distributed:
\begin{equation} \label{eq:predict_error} 
\tilde{r}_{i,t+1} = E_t(\tilde{r}_{i,t+1}) + z_{i, t} = r^e_{i,t} + z_{i,t}
\end{equation} 
where $r^e_{t, i} = E_t(\tilde{r}_{i,t+1})$ is the conditional expected net return, at time $t$, of the next time step; and the error of estimation is $z_{i, t}  \sim N(0,\sigma_{i,t}^2)$. The maximization becomes:
\begin{equation} \label{eq:MEU4} 
\max_{x} E_t\{ (1 + x r^e_{i,t} + x z_{i,t}) ^{1-\gamma_{i,t}}\} = \max_{x} \{\frac{1}{\sqrt{2 \pi \sigma_{i,t}^2}} \int_{-\infty}^{\infty} e^{-\frac{z_{i,t}^2}{2\sigma_{i,t}^2}} f(z_{i,t};x, \gamma_{i,t}) dz_{i,t}\},
\end{equation}
where $f(z_{i,t};x,\gamma_{i,t})= (1 + x r^e_{i,t} + x z_{i,t})^{1-\gamma_{i,t}}$. 
In writing down the above equation we implicitly assumed $(1 + x r^e_{i,t} + x z_{i,t})>0$, which is a necessary requirement for the power function to be a valid utility measure. 
The above integral can be further simplified to
\begin{equation} \label{eq:integral} 
\frac{1}{\sqrt{\pi}} \int_{-\infty}^{\infty} e^{-z_{i,t}^2} f(\sqrt{2} \sigma_{i,t} z_{i,t};x, \gamma_{i,t}) dz_{i,t}
\end{equation}
This Gaussian integral cannot be evaluated analytically, but it can be approximated by an expansion based on the roots of Hermite Polynomial $H^{(n)}(\xi)$ as:
\begin{equation} \label{eq:EU5} 
\int_{-\infty}^{\infty} e^{-z_{i,t}^2} f(\sqrt{2} \sigma_{i,t}z_{i,t}; x, \gamma_{i,t}) dz_{i,t} = \sum_{k=1}^n \lambda_k^{(n)} f(\sqrt{2} \sigma_{i,t} \xi_k^{(n)}; x, \gamma_{i,t})
\end{equation} 
where $\lambda_k^{(n)} (k=1,2,...,n)$ are the coefficients of the summation, and $\xi_k^{(n)}$ are the roots of $n$th Hermite polynomial $H^{(n)}(\xi)$.
Performing the maximization by setting the derivative with respect to $x$ equal to zero, we obtain the following equation:
\begin{equation} \label{eq:MaxEU1} 
\sum_{k=1}^n \lambda_k^{(n)} [1+ ( r^e_{i,t} + \sqrt{2} \sigma_{i,t} \xi_k^{(n)}) x ]^{-\gamma_{i,t}} \times ( r^e_{i,t} + \sqrt{2} \sigma_{i,t} \xi_k^{(n)} ) = 0
\end{equation}

Since $|r^e_{i,t}| \ll 1$, $\xi_k^{(n)} \sim N(1)$, and $\sigma_{i,t} \ll 1$ for the typical time step of one day or shorter, the above can be approximated as
\begin{equation} \label{eq:MaxEU2} 
\sum_{k=1}^n \lambda_k^{(n)} [1 - \gamma_{i,t} (r^e_{i,t} + \sqrt{2} \sigma_{i,t} \xi_k^{(n)}) x ] \times [r^e_{i,t} + \sqrt{2} \sigma_{i,t} \xi_k^{(n)}] = 0.
\end{equation} 
   
Here we consider an approximation with $n=2$. Note that  $\lambda_1^{(2)} = \lambda_2^{(2)}=\lambda$, $\sqrt{2} \xi_1^{(2)}= -\sqrt{2} \xi_2^{(2)} ~(=1)$, the optimal demand of agent $i$ of risky stock can then be obtained as
\begin{equation} \label{eq:demand_i} 
D_{i,t}=x_{i,t} = \frac{ r^e_{i,t}} { \gamma_{i,t} [(r^e_{i,t})^2 + \sigma_{i,t}^2]} \approx \frac{ r^e_{i,t}} { \gamma_{i,t} \sigma_{i,t}^2} = \frac{E_{i,t}[p_{t+1}+d_{t+1}] - (1+r) p_t} {\gamma_{i,t} \sigma_{i,t}^2 (1+r) p_t},
\end{equation}
which is the Eqn. (\ref{eq:demand}).
In writing down the above approximation we assume $(r^e_{i,t})^2 \ll \sigma_{i,t}^2$, which is certainly true when the time step is one day or shorter.

To get a more accurate approximation of the demand function, higher order Hermite polynomial roots are needed in the summation approximation to the integral in Eqn. (\ref{eq:integral}). It can be shown that, the demand function $x_i$ in a higher order approximation is exactly the same as Eqn. (\ref{eq:demand_i}), except for an overall constant factor.

\section{The Baseline Model}
\subsection{Price prediction}
To use the demand and price setting function derived in the previous section, one still need to incorporate each agent's prediction of the payoff at the next time step, $E_t(p_{t+1}+d_{t+1})$.
We assume all agents use the past price information for price forecasting. The simplest way is to calculate a moving average of the available past prices and use it as a proxy of price forecasting for the next time step $t+1$.
Since investors may have different investment horizons and evaluation strategies, they may use different time lags for their calculation of the moving average of past prices, implying that they have heterogeneous memory lengths (Levy \textit{et al}. (1994)).
We consider each agent has his own $M$ sets of predictors so that he can choose the best one for forecasting the price at the next time step.
Each price predictor $E_{i,j}(p_{t+1} + d_{t+1}), (i=1,2,...,N; j=1,2,...,M)$ is made of a moving average of past $L_{i,j}$ prices with a subjective erroneous stochastic adjustment:
\begin{equation} \label{eq:price_forecast} 
E_{i,j,t}(p_{t+1} + d_{t+1}) = MA_{i,j,t} = MA_{i,j,t-1}(1-\frac{1}{L_{i,j}})  + \frac{1}{L_{i,j}} (p_t + d_t) + \varepsilon_{i,j}
\end{equation} 
where $\varepsilon_{i,j} \sim N(0, \sigma_{p+d})$ is Gaussian random variable.

The conditional variance of the estimation, $\sigma^2_{i,t}$, is assumed to update with a moving average of the squared forecast error:
\begin{equation} \label{eq:varianeUpdate} 
\sigma^2_{i,j,t} = (1-\theta) \sigma^2_{i,j,t-1} + \theta [(p_{t} + d_{t}) - E_{i,j,t-1}(p_{t} + d_{t})]^2,
\end{equation}
where $\theta ~(0<\theta \ll 1)$ is a weighting constant. 

\subsection{Dividend process}
The dividend process is assumed to be a random walk:
\begin{equation} \label{eq:dividend} 
d_t = d_{t-1} + r_d + \epsilon_t,
\end{equation}
where $\epsilon_t$ is an i.i.d. Gaussian with zero mean and variance $\sigma_d$; $r_d$ is the average dividend growth rate.
Note that the dividend process in a real stock market may be more complicated than what we assumed here and it may vary from stock to stock. But our results are not sensitive to the choice of a dividend process.

\subsection{The price setting equation of the baseline model}
For heterogeneous agents with fixed constant risk aversion, the demand function Eqn. (\ref{eq:demand}) and price setting Eqn. (\ref{eq:price_t}) and (\ref{eq:price_tau}) can be written as:
\begin{equation} \label{eq:demand_hetero_cra} 
D_{i,t} =\frac{E_{i,t}[p_{t+1} + d_{t+1}] - p_t(r+1)}{\gamma_i \sigma^2_{i,t} (1+r) p_t }
\end{equation}
and
\begin{equation} \label{eq:price_hetero_cra} 
p_t= \frac{\sum_i^N\frac{ E_{i,t}[p_{t+1} ~+~ d_{t+1}]}{\gamma_i \sigma^2_{i,t} (1+r)}} {N + \sum_i^N \frac{1}{\gamma_i \sigma^2_{i,t}}}
\end{equation} 

\begin{equation} \label{eq:price_tau_hetero_cra} 
p_{t+\tau} = \frac{ \sum_i^N \frac{1}{\gamma_i} (\frac{E_{i,t+\tau}[p_{t+\tau+1}+d_{t+\tau+1}]} {\sigma_{i,t+\tau}^2 (1+r) } -  \frac{1} {\sigma_{i,t+\tau}^2} ) }  { \sum_i^N \frac{1}{\gamma_i} (\frac{E_{i,t}[p_{t+1}+d_{t+1}]} {\sigma_{i,t}^2 (1+r) } -  \frac{1} {\sigma_{i,t}^2})} p_t.
\end{equation}

It can be see that the risk aversion indices of the agents play the role  of  weighting factors in the price setting equations.

If the agents are homogeneous in risk aversion, the above can be further simplified to:

\begin{equation} \label{eq:demand_homo_cra} 
D_{i,t} =\frac{E_{i,t}[p_{t+1} + d_{t+1}] - p_t(r+1)}{\gamma \sigma^2_{i,t} (1+r) p_t }
\end{equation}
and

\begin{equation} \label{eq:price_homo_cra} 
p_t= \frac{\sum_i^N\frac{ E_{i,t}[p_{t+1} ~+~ d_{t+1}]}{\sigma^2_{i,t} (1+r)}} {N \gamma + \sum_i^N \frac{1}{\sigma^2_{i,t}}}
\end{equation} 

\begin{equation} \label{eq:price_tau_homo_cra} 
p_{t+\tau} = \frac{ \sum_i^N (\frac{E_{i,t+\tau}[p_{t+\tau+1}+d_{t+\tau+1}]} {\sigma_{i,t+\tau}^2 (1+r) } -  \frac{1} {\sigma_{i,t+\tau}^2} ) }  { \sum_i^N (\frac{E_{i,t}[p_{t+1}+d_{t+1}]} {\sigma_{i,t}^2 (1+r) } -  \frac{1} {\sigma_{i,t}^2})} p_t.
\end{equation}

From the above equations we see that the risk aversion index $\gamma$ only affects the overall level of demand but not its fluctuations. It also does not contribute to the price fluctuations (between the time $t$ and time $t+\tau$).
Thus in the case of homogeneous and constant risk averse agents, the main source of the price fluctuations is from the investors' price forecasting.
The baseline model does not incorporate investor's changing sentiment.
As a consequence, the price fluctuation is expected to be very limited and we will subsequently show that
this is indeed the case. 

\section{Model with Dynamic Risk Aversion}
\subsection{Heterogeneous and dynamic risk aversion}
To extend the baseline model, we allow agents to have heterogeneous risk aversion indices (degrees), which vary with time. This reflects the fact that in a real financial market 
investors have different risk attitudes and the investors' sentiment change with time.
We assume that the risk aversion index of each agent follows an independent bounded random walk with a constant variance $\delta^2$:
\begin{equation} \label{eq:lambda_i_t0} 
\gamma_{i,t} = \gamma_{i,t-1} +\delta z_{i,t} \hspace{5mm}, \gamma_{i,t} \in [\gamma_0, \gamma_u]
\end{equation} 
where $z_{i,t}$ is an i.i.d. Gaussian variable with mean zero and unit variance for agent $i$, $\gamma_0 (>0)$ is the lower boundary, and $\gamma_u (>\gamma_0)$ the upper boundary.
It's easy to relate the value of the index at time $t+\tau$ to the value at time $t$:
\begin{equation} \label{eq:lambda_i_t1} 
\gamma_{i,t+\tau} = \gamma_{i,t} +\delta \sum_{t=1}^{\tau} z_{i,t} = \gamma_{i,t} +\delta S_{i,\tau}
\end{equation} 
where $S_{i, \tau}=\sum_{t=1}^{\tau} z_{i,t}$ is the change of risk aversion index of agent $i$ from time $t$ to time $t+\tau$, which can be either positive or negative.  
It's worthwhile to note that in real markets, the dynamics of investors' risk aversion attitudes may be more complicated than a simple random walk process we assume here. However, simplifying and idealizing of the real situation helps us to stay focused on the main purpose of investigating the impact of investors' fluctuating risk aversion on the price dynamics.

\subsection{Price setting equation with dynamic risk aversion}
Upon substitution of Eqn. (\ref{eq:lambda_i_t1}) into Eqn. (\ref{eq:price_tau}),  we have:
\begin{eqnarray} \label{eq:price_tau_dra0} 
p_{t+\tau}= \frac { \sum_i^N \frac{1}{(\gamma_{i,t} +\delta S_{i, \tau})} (\frac{E_{i,t+\tau} (p_{t+\tau+1}+d_{t+\tau+1})}{ \sigma^2_{i,t+\tau} (1+r) }  -  \frac{1}{\sigma^2_{i,t+\tau}}) } {\sum_i^N \frac{1}{\gamma_{i,t}} (\frac{E_{i,t} (p_{t+1}+d_{t+1})}{ \sigma^2_{i,t} (1+r) }  -  \frac{1}{\sigma^2_{i,t}}) } p_t
\end{eqnarray}
Comparing Eqn.(\ref{eq:price_tau_dra0}) to Eqn. (\ref{eq:price_tau_hetero_cra}) ($\gamma_{i,t}=\gamma_{i,t+\tau}=\gamma_i$) for the case of fixed constant risk aversion, we see that there is an extra term, $\delta S_{i,\tau}$, in the price setting equation in the case of DRA. 
Since $S_{i,\tau}$ can be either positive or negative and its value changes with time, $\gamma_{i,t} +\delta S_{i,\tau}$ deviates from $\gamma_{i,t}$ and fluctuates with time.
This fluctuating weighting factor (representing agent's fluctuating risk aversion) acts like an `amplifier'' of the price deviation induced by the error in agents' price estimation, and therefore results in excess price fluctuation.
$|S_{i,\tau}| \sim N(0, \sqrt{\tau})$,  for $\sqrt{\tau} \delta \gg 1$, $\gamma_{i,t+\tau} (\gamma_{i,t} +\delta S_{i,\tau})$ and $\gamma_{i,t}$ can differ substantially, resulting in a large deviation of $p_{t+\tau}$ from $p_t$.
Our numerical results, presented in the next session, clearly show that it's this risk aversion dynamics that gives rise to the excessive price fluctuations and the associated volatility clustering.

\subsection{The range of DRA indices}
We now examine the range of possible relative risk aversion indices.  
The choice of the range is important for modeling investors' decision-making; it has big impact on the price dynamics, as it directly affects investors' demand of the risky asset.  The lower the index, the less risk-averse the investor is (thus the higher the demand of risky asset); and vice versa. Thus the risk-aversion attitude has great impact on the price dynamics through its influence on the demand.  
Some empirical and experimental studies reported that for a ``typical'' investor, the value of the risk-aversion index $\gamma$ is in the range of 0-2 (Mehra and Prescott (1985), Friend and Blume (1975), Levy, Levy and Solomon (2000)). 
Mehra and Prescott (1985) used a value of risk aversion index with an upper limit of $10$  in their treatment of the issue of the ``Equity Premium Puzzle''.
However, to ``explain'' the ``Equity Premium Puzzle'' of NYSE over 50 years of U.S. postwar period, one needs a relative risk aversion index of \textit{250} if a consumption-based model is used(Cochrane (2005))! These empirical results show that it is better to model the risk aversion with a range of indices, instead of a fixed value. The range we specified consists of an upper bound and a lower bound for the random walk describing DRA indices.

\section{The Simulation Results and Analysis}
\subsection{The setup}
In our simulation we choose the number of agents $N=100$, the number of predictors each agent has, $M=2$.
Setting different number of agents produces similar results.
The initial risk aversion indices $\gamma_{i,0}$ are all set to 1.0 for the baseline model and are set to $\gamma_{i,0} \in [0.2, 4]$ for the model with DRA.
The bounds for the index of DRA are  $\gamma_{i,t} \in [10^{-5}, 20]$,
the risk-free interest rate is $r=5\%$, the dividend growth rate is $r_d = 2\%$,
and the weighting coefficients for the variance of estimation is $\theta = 1/250$.
The lags used in the price estimators are $L_{i,j} \in [2,250]$, and we set $\sigma_{p+d}$=1\% for all agents.

\subsection{Simulation price and trading volume}
Let's first take a look at how excess price fluctuations emerge from a dynamic risk aversion process. Fig.1 shows the simulation time series of the price and the trading volume generated from the model with fixed constant risk aversion (CRA) and the model with dynamic risk aversion (DRA).

From the figure we see clearly that the DRA leads to increased fluctuations in both the price and the trading volume. To have both qualitative and quantitative picture of the impact of the DRA on the price dynamics, we examine the key stylized facts in the following subsections. 

\subsection{Autocorrelation function}
One of the stylized facts observed in real financial data is that their autocorrelation functions (ACF) usually start with a low value (from $\rho_1$) and decay very slowly with increase of time step for the squared or absolute-valued returns.
For an almost-Gaussian process, the values of its ACFs for absolute-valued return are close to zero and independent of the time steps.
In Fig.2, we compare the ACFs for absolute-valued returns for the series generated by our baseline model (with CRA) and the DRA model, the series of real data (DJIA and SP500 Index), and Gaussian process.

The figure clearly shows that while the ACFs generated from our baseline model (CRA) is close to that of Gaussian process (Gauss), the results generated from our DRA model are very close to the real financial data (DJIA, SP500).

\subsection{Excess volatility}
The second key stylized facts we examine is the excess volatility (or fat-tails) of returns, which measures the price fluctuation of real financial series. In Fig.3 we plot the distributions of returns (in different time-steps) from our baseline model and the DRA model with the parameters set according to a normal market condition. For comparison, we also plot the return distribution of DJIA and the return distribution generated by a simple Gaussian process.

These plots show that, for all different time periods, the return distributions from our baseline model are very close to those of the Gaussian process; in contrast, the results from the model with DRA are close to the real DJIA data.
In the context of our model, it is clear that the dynamic risk aversion leads to excess volatility or a fat-tail in the return distribution, which is one of the most important characterization of real financial time series (Mandelbrot (1963), Fama (1965)).

To further examine the fat tail of the distribution, we plot, in Fig.4, the Kurtosis as a function of the squre root of variance of risk aversion, $\delta$.
From these plots, we see that the risk aversion dynamics can change the return distribution significantly
from a Gaussian distribution (which has $K$=3.0). In addition, the smaller the lag $\tau$, the larger the Kurtosis generated;
this is consistent with the empirical observations in real financial data. These values of the Kurtosis, together with the standard deviation and skewness are listed in Table 1.
Note that the statistics generated from our DRA model are quite close to that from DJIA daily data.

\subsection{Volatility Clustering}
Volatility clustering is another important characteristics of financial time series. 
Here we use the conditional probability measure developed recently by Chen \textit{et al.} (2005) to examine the volatility clustering.
The method uses the return distribution conditional on the absolute return in the previous period to describe a functional relation between the variance of the current return and the absolute return in the previous period.
If the volatility in asset returns is clustered, it will be proportional to the volatility in the previous period, the proportionality constant reflects the strength of volatility clustering. 

Fig.5 shows how the current volatility depends on the volatility of the previous period for a random walk process, the baseline market model, DRA model, and DJIA daily data.
From the figures we see clearly that the baseline market model with fixed constant risk aversion (CRA) produces very low volatility clustering (the curve is flat if there is no volatility clustering, such as the case with the random walk model (Gauss)). 
In contrast, the volatility clustering from the DRA model is significantly higher, and it is very close to the one generated from DJIA daily prices.
The plots also show that, for our DRA model, the smaller the time period (step) of the return, the stronger the volatility clustering; and vice versa; this is consistent with the empirical observations of real market data.

\section{The SFI market model with dynamic risk aversion}

\subsection{Brief introduction to SFI market}
To test the impact of DRA on other baseline model, we use the well-known Santa Fe model of artificial market (Arthur \textit{et al.} (1997), LeBaron, Arthur and Palmer (1999)). We first give a brief summary of the SFI market below.
In the SFI market a constant absolute risk aversion (CARA) utility function ($U(c_t,\gamma)=-e^{c_t\gamma}$) is assumed for each agent, the demand and price setting equations in fact have similar forms as in Eqn.~(\ref{eq:price_t}) and (\ref{eq:price_tau}).

The dividend process is assumed to be an AR(1) process:
\begin{equation} \label{eq:dividend} 
d_t = \bar{d} + \rho (d_t - \bar{d}) + \varepsilon _t
\end{equation}
where $\epsilon_t$ is an i.i.d. Gaussian with zero mean and variance $\sigma_e$
    
The SFI market assumes that each of the $N$ (=25) agents at any time possesses $M$ (=100) linear predictors and uses those that best matches the current market state and have recently proved most accurate. 
Each predictor is a linear regressor of the previous price and dividend, $E(p_{t+1}+d_{t+1}) = a (p_t +d_t) + b$; it uses a market state ``recognizer'' vector consisting of $J$ (=12) elements, each taking a value of either $0$, $1$ or \#(match any market states).
The market status is described by a state vector consisting of $J$ binary elements, each taking value of either $1$ (its specified market condition exists) or $0$ (the market condition does not exist).
The elements of market state can represent any important market discriminative information, including macro-/micro- economic environment, summary of fundamentals, and market temporal trends, etc.
At each time step, only those predictors which match all their $J$ elements to the  corresponding $J$ elements of market status are eligible to be used and are called ``active'' predictors.

The variance of estimation $\sigma^2_{i,t}$ for each agent is assumed to update with a moving average of squared forecast error, defined in Eqn. (\ref{eq:varianeUpdate}).

Agents learn to improve their performance by discarding the worst (20\%) predictors and developing new predictors via a genetic algorithm.
This ensures the market some dynamics.

In the SFM, the market conditions are specified as:

1-6 elements represent ``current price $\times$ interest rate / dividend $>$ 0.25, 0.5, 0.75, 0.875, 1.0, 1.125.

7-10 elements describe ``Current price $>$ 5-period moving average of past prices (5-period MA), 10-period MA, 100-period MA, 500-period MA.

11th element always 1;

12th element always 0;

The regressor's parameters $\{a,b\}$ are set to be randomly and uniformly distributed within the ranges: $a\in (0.7, 1.2)$ and $b\in (-10, 19.002)$.

The risk-free interest rate is set to 5\%, and for the dividend process: the auto-regression coefficient $\rho$ = 0.95, $\bar{d}=10$, and $\sigma_e = 0.0745$.

The weighting coefficients for the variance of estimation, $\theta = 1/75, 1/150$ for faster and slower learning respectively.
The genetic algorithm is invoked (on average) every $T_e=250$ periods (faster learning) and 1000 periods (slower learning). For more detailed justification for choosing the parameter values, see LeBaron \textit{et al.}(1999).

With these setups, we have checked that the simulation stock price time series and its statistical properties generated are in fact similar to those of our baseline model.

\subsection{Numerical results of SFI market model with DRA}
The dynamics of risk aversion can be similarly incorporated into the SFI market model.
Fig.6 plots the simulation price time series for different DRA variance $\delta^2$.
These plots show clearly the impact on price fluctuations from the DRA.
To have a quantitative picture on how the excess volatility emerges from the DRA, we plot in Fig.7 the functional relation of the Kurtosis vs variable $\delta$. The results are very similar to those plotted in Fig. 4, which were obtained with our much simpler baseline model.

We have checked other key features and found that the SFI-DRA model gives the similar results as our DRA model which is based on a much simpler baseline model.
This suggests that DRA is the key mechanism for the emergence of the key stylized facts, and the impact of DRA does not depend on the structures of the baseline models.
Therefore the price impact of investors' DRA we have studied is generic.

\section{Summary}
We have presented a simple multi-agent model of a financial market which incorporates the dynamics of risk aversions. We assume that the index of DRA follows a simple independent bounded random walk with a constant variance $\delta^2$.
We demonstrate that such dynamics is directly responsible for excess volatility and the associated volatility
clustering. We compare the numerical results from our model with the results obtained by analyzing the DJIA daily data and show that the simulation data reproduce most of the ``stylized'' facts, such as excess volatility (measured by fat tail and high peak of return distribution), volatility clustering measured by conditional return distribution.
We have also tested the DRA on the Santa Fe market model and obtain similar results.
This suggests that the impact of DRA among heterogeneous agents we introduced here does not depend on the structure of the particular baseline model used. The degree of excess volatility is essentially controlled by the parameter $\delta$. Thus $\delta$ can be used as a key market sentiment parameter, in conjunction with the other market indicators such as average return $r$ and the average volatility $\sigma_0$, to characterize the financial market. We hope that our results presented here will provide new insights into the dynamics of asset price fluctuations governed by investors' fluctuating sentiments.
\\
\\
 
\begin{Large} \bf{Acknowledgment}
\end{Large}
\\
\\
Baosheng Yuan is deeply grateful to Blake LeBaron for his very helpful and illuminating suggestions at several points in the research.

\begin{figure}[h]
\begin{center}
\includegraphics[height=12cm, width=12cm]{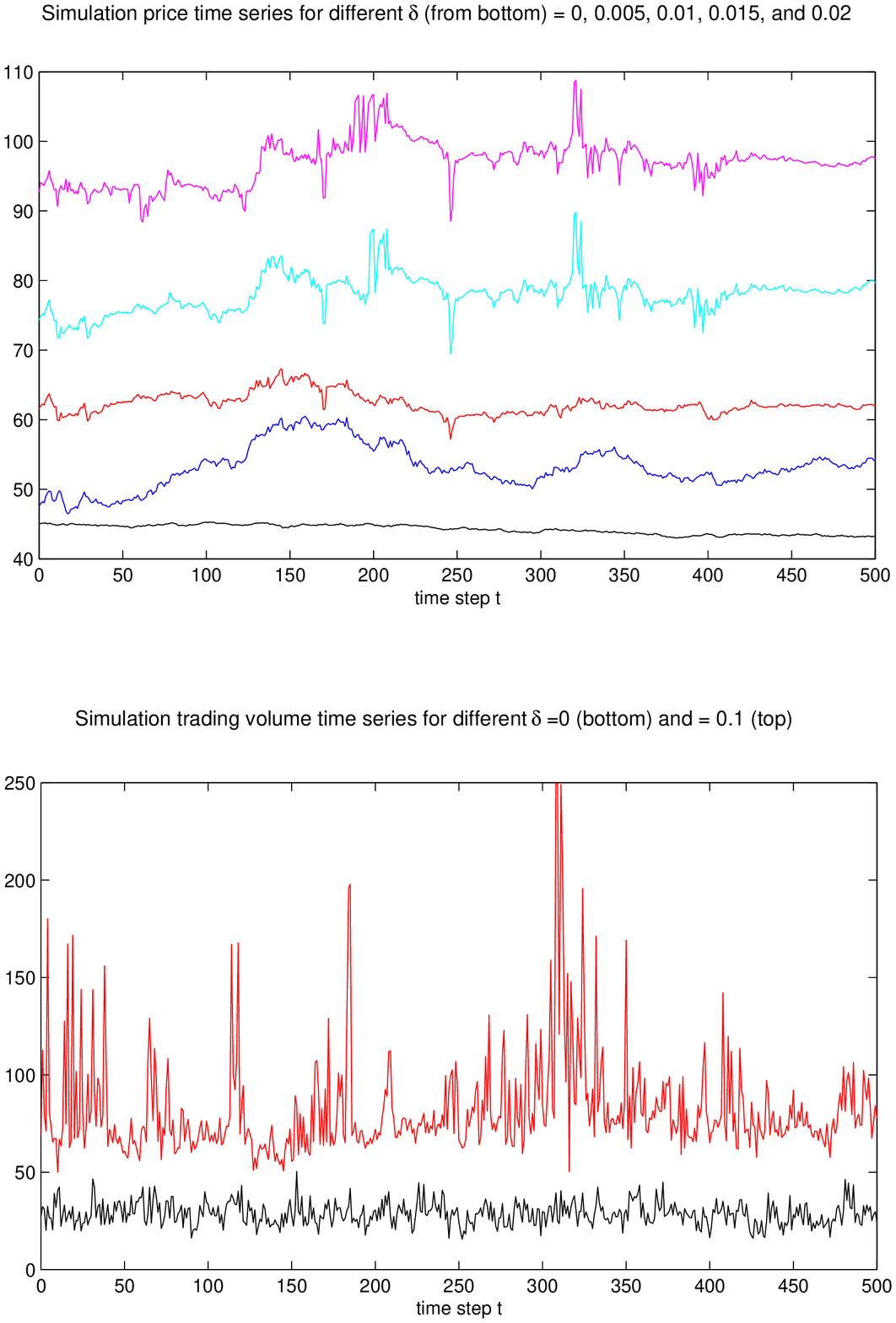}
\caption{\footnotesize{The time series of price and trading volume from the models with constant ($\delta = 0$) and dynamic ($\delta \neq 0$) risk aversion. For the sake of clarity, the time series with different $\delta$s were vertically shifted, e.g., the trading volume for $\delta$ =0.01 was upward shifted by 50.}}
\end{center}
\end{figure}

\begin{figure}[h]
\begin{center}\includegraphics[height=8cm, width=14cm]{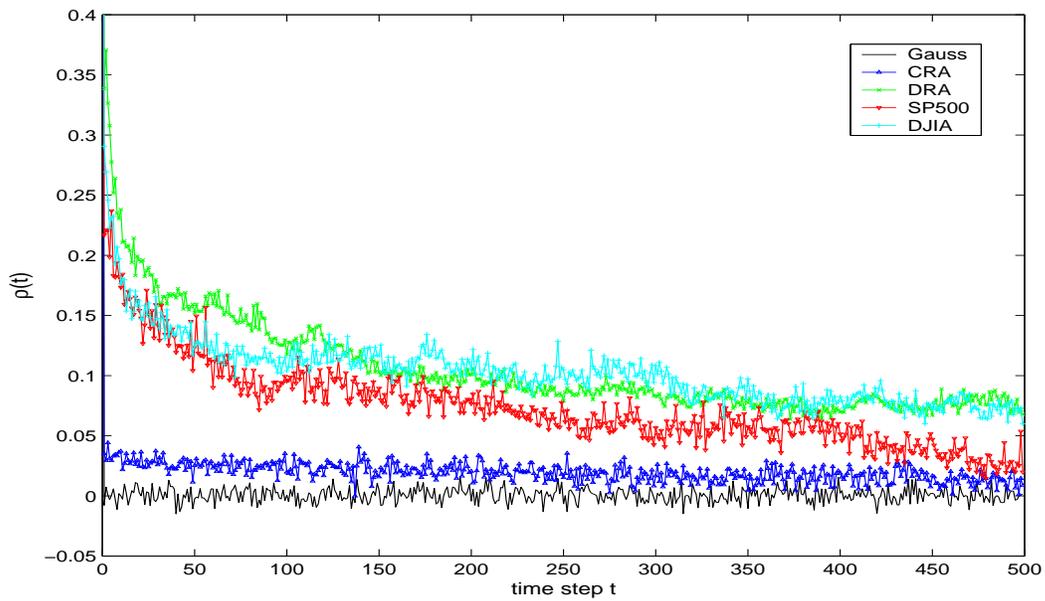}
\caption{ \footnotesize{ACFs of different time steps for return series of: the baseline model (CRA), the DRA model, DJIA Index, SP500 Index, and random walk process (Gauss).}} \end{center}
\end{figure}

\begin{figure}[h]
\begin{center}\includegraphics[height=12cm, width=14cm]{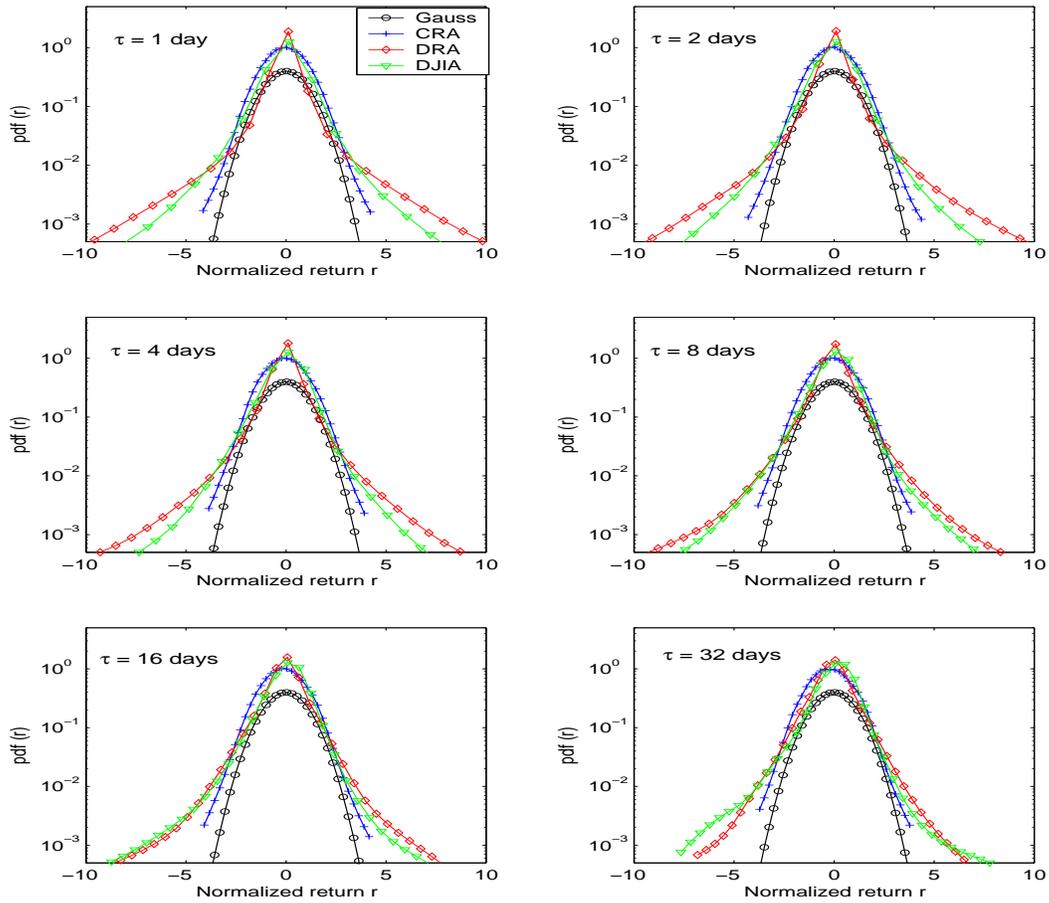}
\caption{ \footnotesize{Return distributions for the model with constant risk aversion (CRA) ($\delta$=0) and the DRA model ($\delta$=0.01), Gaussian process (Gauss) and real data of DJIA.}} \end{center}
\end{figure}

\begin{figure}[h]
\begin{center}\includegraphics[height=6cm, width=12cm]{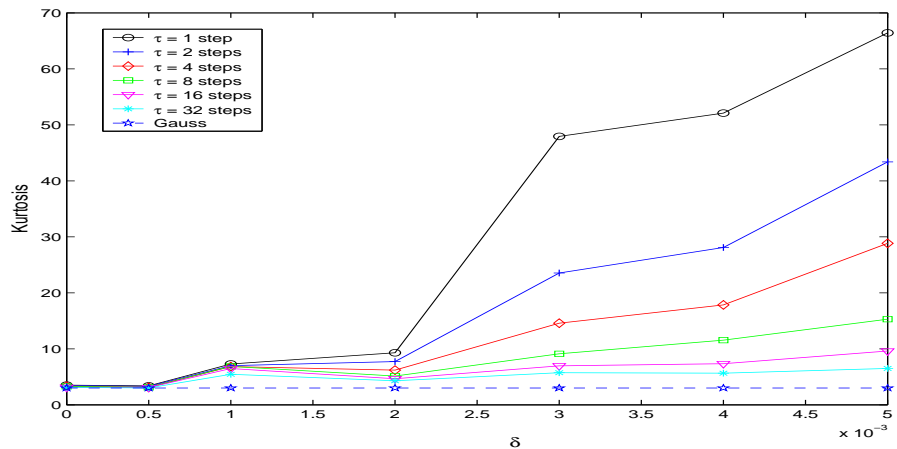}
\caption{ \footnotesize{Kurtosis of simulation price series for different variables $\delta$ of the DRA index.}} \end{center}
\end{figure}

\begin{table}[h] \begin{center}
\caption{Some statistics for DRA model ($\delta$=0.01) and DJIA}
\vspace{0.8 ex}
\begin{tabular}{|c c c c|}
\hline
Lag ($\tau$) &  S.D. & Skewness & Kurtosis \\
\hline
& & Model $\delta$ = 0.01 &\\
\hline
{\scriptsize 1}  &{\scriptsize 0.015004}  &{\scriptsize -0.454483}  &{\scriptsize 48.406593}\\
{\scriptsize 2}  &{\scriptsize 0.018313}  &{\scriptsize -0.372956}  &{\scriptsize 35.233093}\\
{\scriptsize 4}  &{\scriptsize 0.021685}   &{\scriptsize -0.460952}  &{\scriptsize 26.051126}\\
{\scriptsize 8}  &{\scriptsize 0.026179}   &{\scriptsize -0.492277}  &{\scriptsize 17.309658}\\
{\scriptsize 16}  &{\scriptsize 0.031840}   &{\scriptsize -0.424137}  &{\scriptsize 11.376849}\\
{\scriptsize 32}  &{\scriptsize 0.039121}   &{\scriptsize -0.327192}  &{\scriptsize 7.444087}\\
\hline
& & DJIA  &\\
\hline
{\scriptsize 1} &{\scriptsize 0.010876} &{\scriptsize  -1.190497}  &{\scriptsize  40.2330} \\
{\scriptsize 2} &{\scriptsize 0.015710} &{\scriptsize  -1.100639}  &{\scriptsize  29.9739} \\
{\scriptsize 4} &{\scriptsize 0.022213} &{\scriptsize  -0.950496}  &{\scriptsize  18.9213} \\
{\scriptsize 8} &{\scriptsize 0.032103} &{\scriptsize  -0.989078}  &{\scriptsize  13.6606} \\
{\scriptsize 16} &{\scriptsize 0.046404} &{\scriptsize  -0.995792}  &{\scriptsize  12.0792} \\
{\scriptsize 32} &{\scriptsize 0.066602} &{\scriptsize  -0.706621}  &{\scriptsize  9.5892} \\
\hline
\end{tabular} \end{center} \end{table}

\begin{figure}[h]
\begin{center}\includegraphics[height=14cm, width=14cm]{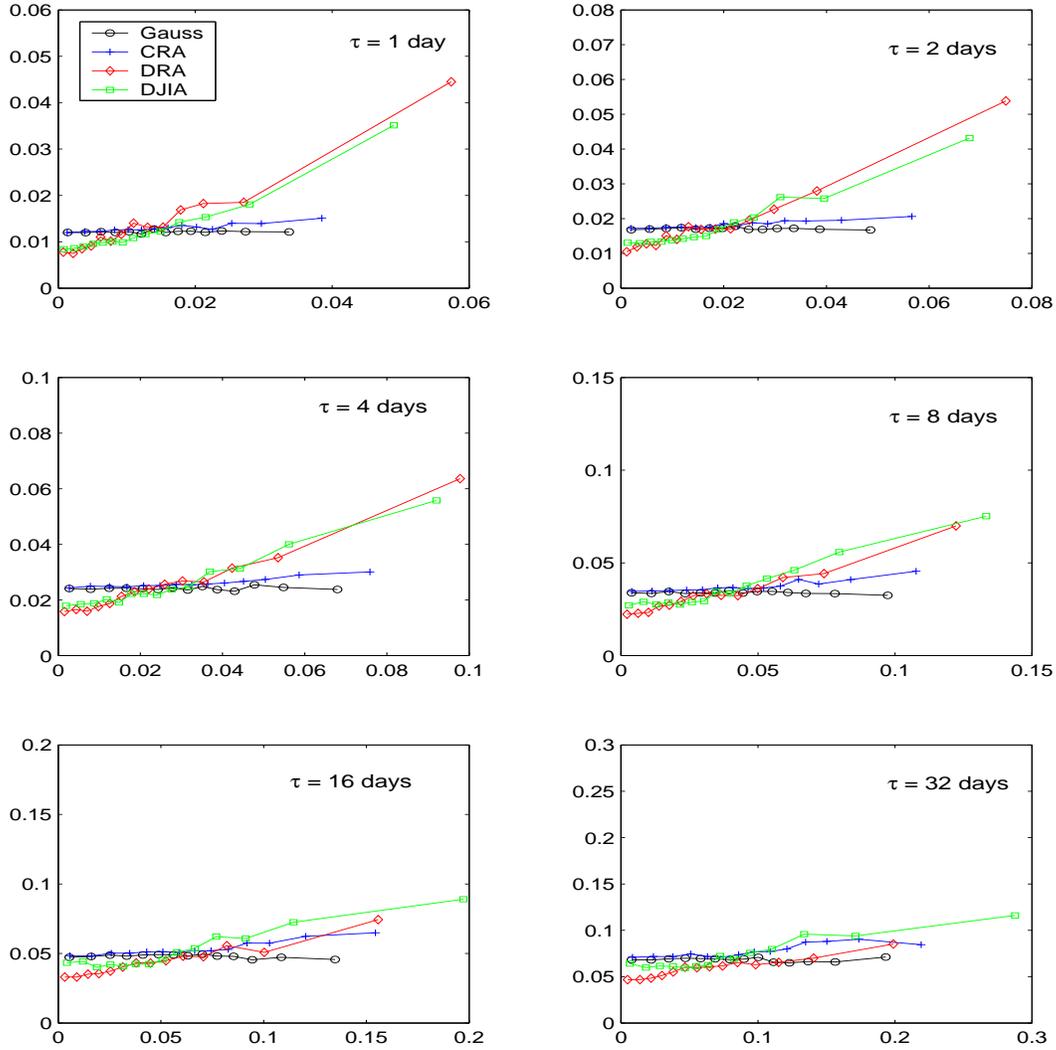}
\caption{ \footnotesize{The volatility clustering measured by the standard deviation of current return vs the absolute return of previous period.}} \end{center}
\end{figure}

\begin{figure}[ht]
\begin{center}
\includegraphics[height=6cm, width=12cm]{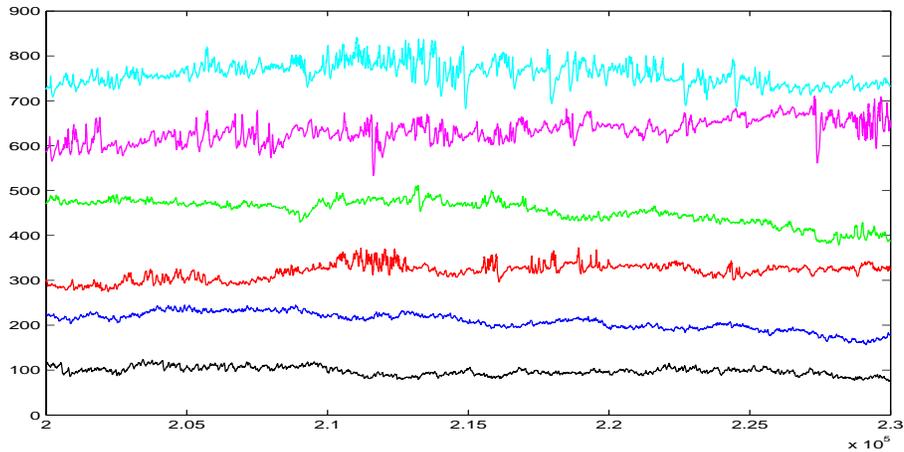}
\caption{ \footnotesize{Simulation price series for different variances of DRA processes with Santa Fe model.  The model parameters are: $\theta=1/75$, $T_e=250$. The prices for different $\delta$s have been vertically shifted to make the comparison clearer. From the bottom to top, the price series are respectively for $\delta$ = 0, 0.01, 0.02, 0.03, 0.04 and 0.05.}} \end{center}
\end{figure}

\begin{figure}[ht]
\begin{center}
\includegraphics[height=6cm, width=12cm]{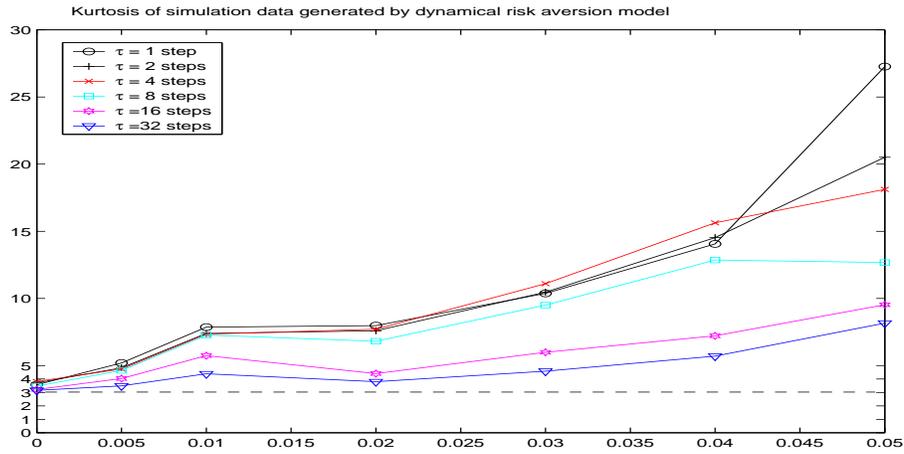}
\caption{ \footnotesize{Kurtosis of simulation price series for different variables, $\delta$, of DRA processes with Santa Fe model.
The model parameters are: $\theta=1/75$, $T_e=250$;
The data were obtained by averaging over 50 independent runs.}}
\end{center}
\end{figure}

\end{document}